\documentclass{article}

\usepackage{arxiv}

\usepackage[utf8]{inputenc} % allow utf-8 input
\usepackage[T1]{fontenc}    % use 8-bit T1 fonts
\usepackage{hyperref}       % hyperlinks
\usepackage{url}            % simple URL typesetting
\usepackage{booktabs}       % professional-quality tables
\usepackage{amsfonts}       % blackboard math symbols
\usepackage{nicefrac}       % compact symbols for 1/2, etc.
\usepackage{microtype}      % microtypography
\usepackage{lipsum}
\usepackage{graphicx}
\usepackage{float}
\usepackage{subfigure}
\usepackage{amsmath}
\graphicspath{ {./images/} }

\title{Counterfactual time series analysis for the air pollution during the outbreak of COVID-19 in Wuhan}

\author{
	Chenran Weng \\
	University of Science and Technology of China\\
	Hefei, China \\
	\texttt{wcr2020flavia@mail.ustc.edu.cn} \\
	\And
	Weiyi He \\
	University of Science and Technology of China\\
	Hefei, China \\
	\texttt{hwyii@mail.ustc.edu.cn} \\
	\And
	Wenjing Zhao \\
	University of Science and Technology of China\\
	Hefei, China \\
	\texttt{zwj0612@mail.ustc.edu.cn} \\
}

\begin{document}
	\maketitle
	\begin{abstract}
		Environmental issues are becoming one of the main topics of concern for society, and the quality of air is closely linked to people's lives. Previous studies have examined the effects of abrupt interventions on changes in air pollution. For example, researchers used an interrupted time series design to quantify the impact of the 1990 Dublin coal ban; and a regression discontinuity to determine the arbitrary spatial impact of the Huaihe River policy in China. An important feature of each of these studies is that they investigated abrupt and localized changes over relatively short time spans (the Dublin coal ban) and spatial scales (the Huaihe policy). Due to the abrupt nature of these interventions, defining a hypothetical experiment in these studies is straightforward. In response to the novel coronavirus outbreak, China implemented "the largest quarantine in human history" in Wuhan on January 23, 2020. Similar measures were implemented in other Chinese cities. Since then, the movement of people and associated production and consumption activities have been significantly reduced. This provides us with an unprecedented opportunity to estimate the changes in air pollution brought about by this sudden "silent" move. We speculate that the initiative will lead to a significant reduction in regional air pollution. Thus, we performed \cite{dey2021counterfactual} counterfactual time series analysis on Wuhan air quality data from 2017-2022 based on three models, SARIMA, LSTM and \cite{chen2016xgboost}XGBOOST, and compared the excellence of different models. Finally, we conclude that "silent" measures will significantly reduce air pollution. Using this conclusion to further investigate the extent of air pollution reduction will help the country to better designate environmental policies. 
	\end{abstract}

	% keywords can be removed
	\keywords{Counterfactual time series analysis, SARIMA, LSTM, XGBoost}

	\section{Introduction}
	The outbreak of COVID-19, declared a pandemic by the WHO, has become a major crisis for global health and socioeconomic generation. To level the epidemic curve and effectively protect public health and safety, the government has issued unprecedented embargo measures. Based on previous experimental and observational studies, the embargo measures have been highly successful in breaking the chain of transmission of COVID-19 and preventing infection in susceptible populations. In turn, the strict measures have led to a marked and rapid improvement in the environment. As a result of the lockdown restrictions, production, industrial processes and construction operations in small and medium-sized enterprises were suspended. Transportation activities also followed a minimalist pattern, including only the transportation of basic household and medical supplies. Whereas traﬀic emissions are a significant source of air pollution, this situation provides a unique opportunity to study the effects of the COVID-19 blockade on traﬀic-related air pollution. In the course of our study, we mainly adopt a counterfactual prediction approach, based on SARIMA model, LSTM model, and XGBoost model, respectively, to fit the air pollution data of Wuhan from 2017-2019 to predict the air pollution level from January 1, 2020.

	\section{Data Source}
	\label{sec:headings}
	Data from Qingyue Data (www.epmap.org) - Gas - National Control Air - National Control City Daily. Search for data from 2017-01-01 to 2023-01-01 and download to get the raw data.
	
	\section{Methods}
	\label{sec:others}
	\subsection{Counterfactual time series analysis}
	Counterfactual time series analysis is a method of studying what would have happened in the past if a certain event or decision had been different. It allows to simulate different scenarios and understand how they would have affected the outcome. In our literature, we predict the concentration of the pollutants, considering them as the concentration without the pandemic, and then compare them with the observed pollutant values.
	\subsection{SARIMA Model}
	SARIMA (Seasonal Autoregressive Integrated Moving Average), an extension of the ARIMA model, is a type of time series model that is used to analyze and forecast data that exhibits both seasonality and autocorrelation. SARIMA adds 3 new parameters to the ARIMA model: the seasonal period, the number of seasonal differences, and the number of seasonal moving average terms, allowing the model to account for patterns in the data that repeat at regular intervals.
	
	The SARIMA(p,d,q)(P,D,Q)s model is given by the following equation:
	$$
	ARIMA(p,d,q)+Seasonality(P,D,Q)s
	$$
	where\\
	$p$ is the order of the autoregressive term\\
	$d$ is the degree of differencing\\
	$q$ is the order of the moving average term\\
	$P$ is the order of the seasonal autoregressive term\\
	$D$ is the degree of seasonal differencing\\
	$Q$ is the order of the seasonal moving average term\\
	$s$ is the seasonal period, or the number of time steps in each season\\
	
	The basis of the SARIMA model is a linear regression of a response variable $Y_t$ at time t against the past values $(Y_{t-1}, Y_{t-2}, ...)$ of Y and the past forecast errors $(\epsilon_{t-1}, \epsilon_{t-2}, ...)$. They have the advantage of accounting for the time trend, seasonality, confounders, and residual autocorrelation.
	\subsection{LSTM Model}
	The deep learning model LSTM is an extension of RNN, which can solve the gradient disappearance problem very concisely. LSTM model essentially extends the memory of RNN, enabling them to maintain long-term dependence on learning input. This memory expansion owns the ability to remember information for a long time. LSTM memory is called a "gated" cell, and the inspiration for the word "gate" comes from the ability to retain or ignore memory information\cite{siami2019performance}. The LSTM model captures important features from the input and saves the information for a long time. Deleting or retaining information is based on the weight value assigned to the information during training. Therefore, the LSTM model learns which information is worth preserving or deleting.
	
	Generally, the LSTM model consists of three gates: the forget gate, the input gate, and the output gate. The forget gate decides to retain or delete the existing information. The input gate specifies the extent to which the new information is added to the memory. And the output gate controls whether the existing value in the current cell is helpful for output.
	
	\begin{figure}[!htb]
		\centering
		\includegraphics[scale=0.1]{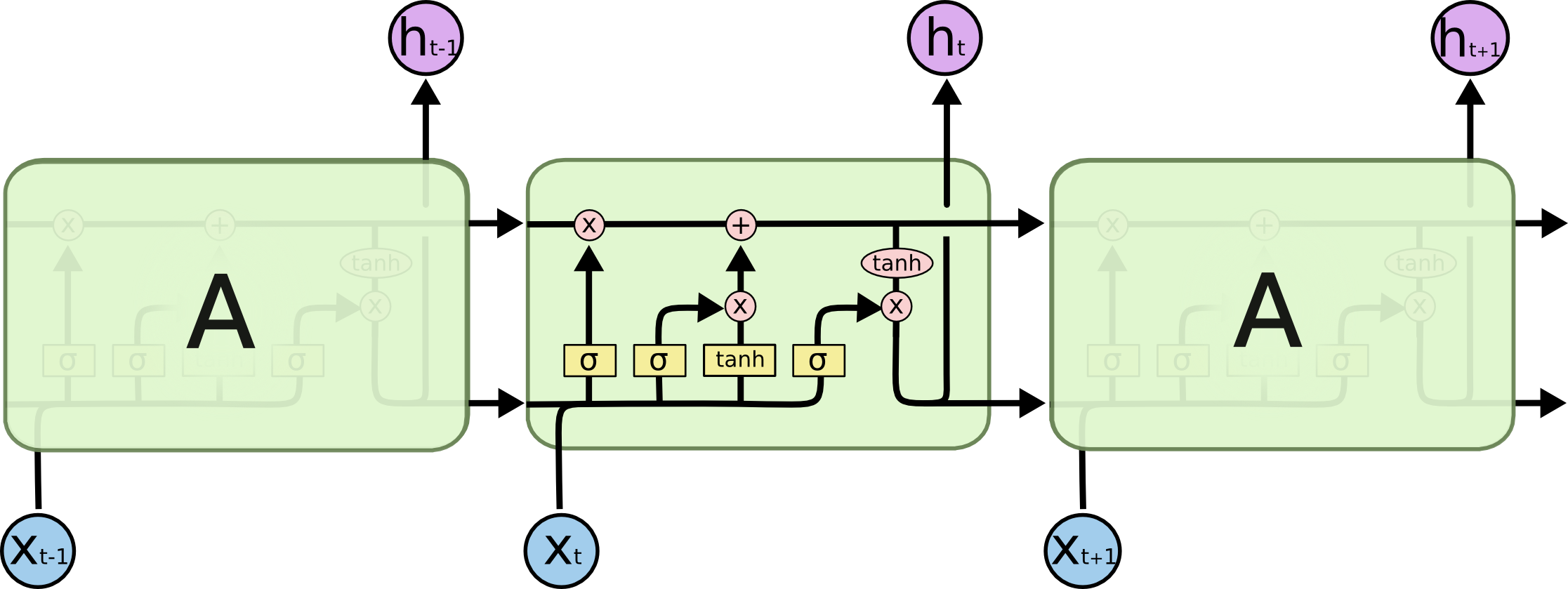}
	\end{figure}

	\paragraph{Forget Gate}
	This gate uses the value of $h_{t-1}$ and $x_t$ to determine whether the information should be deleted from the memory with the help of the sigmoid function. The output of this gate is $f_t \in [0,1]$ where 0 means completely removing the learned value, and 1 means keeping the whole value:
	$$
	f_t = \sigma(W_f\cdot[h_{t-1},x_t] + b_f)
	$$
	where $W_f$ is weight martix and $b_f$ is the bias value. 
	
	\paragraph{Input Gate}
	This gate consists of a sigmoid layer and a tanh layer to determine whether the information should be added to the memory. The output of these two layers is calculated as:
	$$
	\operatorname{Sigmoid}: i_t = \sigma(W_i\cdot[h_{t-1},x_t] + b_i)
	$$
	$$
	\operatorname{Tanh}: \tilde{C_t} = tanh(W_C \cdot[h_{t-1},x_t] + b_C)
	$$
	Then the LSTM memory would be renewed with the function:
	$$
	C_t = f_t \ast C_{t-1} + i_t \ast \tilde{C_t}
	$$
	where $C_{t-1}$ represents the old memory and $\tilde{C_t}$ means the added memory.
	
	\paragraph{Output Gate}
	The first step of the output gate still uses the sigmoid layer to calculate $o_t$, and decides which part needs to be put out:
	$$
	o_t = \sigma(W_o\cdot[h_{t-1},x_t] + b_o)
	$$
	The second step is to deal values with the tanh layer in order to put out $h_t \in [-1,1]$:
	$$
	h_t = o_t \ast \operatorname{tanh}(C_t)
	$$

	\subsection{XGBoost Model}
	The full name of XGBoost is eXtreme Gradient Boosting. If the weak classifier generation in each step of the boosting algorithm is based on the gradient direction of the loss function, it is called gradient boosting. The XGBoost algorithm is a stepwise forward additive model, except that a coefficient no longer needs to be computed after generating the weak learner in each iteration. XGBoost is an additive operator consisting of k base models
	$$\hat{y_i}=\sum_{t=1}^kf_t(x_i)$$
	where $f_k$ is the kth base model and $ \hat{y_i}$ is the predicted value of the i-th sample. Then the loss function can be represented by the predicted value $ \hat{y_i}$ and the true value $y_i$ as follows.
	$$L=\sum_{i=1}^nl(y_i,\hat{y_i})$$
	where n is the number of samples.
	The XGboost model is constructed in the following steps.

	\paragraph{Define objective function}
	
	The prediction accuracy of the model is determined by both the bias and the variance. The loss function represents the bias of the model, and a simpler model is needed if you want the variance to be small, so the objective function ultimately consists of the loss function L and the regular term $\Omega$ that suppresses the complexity of the model. The objective function is as follows.
	$$Obj = \sum _ { i = 1 } ^ { n } l ( \hat { y } _ { i } , y _ { i } ) + \sum _ { t = 1 } ^ { k } \Omega ( f _ { t } )$$
	where $\Omega ( f _ { t } )$ is the regularization term
	$$\Omega ( f _ { t } ) = \gamma T _ { t } + \frac { 1 } { 2 } \lambda \sum _ { j = 1 } ^ { T } w _ { j } ^ { 2 }$$
	The preceding $T _ { t }$ is the number of leaf nodes, $w _ { j }$ denotes the node weights on leaf j, and $\gamma$, $\lambda$ are the pre-given hyperparameters. With the introduction of regularization, the algorithm selects a simple model with good performance, and the regularization term is only used to suppress overfitting of the weak classifier $f_{i}(x)$ in each iteration and is not involved in the integration of the final model.
	We know that,boosting model is forward additive, taking the tth step model as an example, the prediction of the model for the i-th sample $x_i$ is
	$$\hat { y } _ { i } ^ { t } = \hat { y } _ { i } ^ { t - 1 } + f _ { t } ( x _ { i } )$$
	where $\hat { y } _ { i } ^ { t - 1 }$ is the predicted value given by the model at step t-1,which is a known constant, and $f_{t}(x_i)$ is the new model that needs to be added, so substituting this into the above, it can be further reduced to
	$$\left. \begin{array} { l }{ O b j ^ { ( t ) } = \sum _ { i = 1 } ^ { n } l ( y _ { i } , \hat { y } _ { i } ^ { t } ) + \sum _ { i = 1 } ^ { t } \Omega ( f _ { i } ) } \\ { = \sum _ { i = 1 } ^ { n } l ( y _ { i } , \ hat { y } _ { i } ^ { t - 1 } + f _ { t } ( x _ { i } ) ) + \sum _ { i = 1 } ^ { t } \Omega ( f _ { i } ) } \end{array} \right.$$
	
	The above is the objective function of XGBoost. Optimizing this objective function is actually equivalent to solving for the current $f_{t}(x_i)$.
	\paragraph{Taylor's simplification of the objective function}
	
	According to the Taylor formula, we expand the function $f ( x + \Delta (x) )$ in second order at the point x to obtain
	$$f ( x + \Delta x ) \approx f ( x ) + f ^ { \prime } ( x ) \Delta x + \frac { 1 } { 2 } f ^ { \prime \prime } ( x ) \Delta x ^ { 2 }$$
	Considering $\hat { y } _ { i } ^ { t - 1 } + f _ { t } ( x _ { i } ) )$ in the $l ( y _ { i } , \hat { y } _ { i } ^ { t - 1 } ) $ of objective function as x and $f_{t}(x_i)$ as $\Delta (x)$, then the objective function can be written as
	$$\left. \begin{array} { l }{O b j ^ { ( t ) } \approx \sum _ { i = 1 } ^ { n } [ l ( y _ { i } , \hat { y } _ { i } ^ { t - 1 } ) + g _ { i } f _ { t } ( x _ { i } ) + \frac { 1 } { 2 } h _ { i } f _ { t } ^ { 2 } ( x _ { i })] + \sum _ { i = 1 } ^ { t } \Omega ( f _ { i } )}\end{array} \right.$$ 
	where $g_i$ is the first order derivative of the loss function l and $h_i$ is the second order derivative of the loss function l. Note that the derivative here is the derivative of $ \hat { y } _ { i } ^ { t - 1 }$.
	$$g _ { i } = \frac { \partial L ( y _ { i } , \hat { y } _ { i } ^ { ( t - 1 ) } ) } { \partial \hat { y } _ { i } ^ { ( t - 1 ) } }  , h _ { i } = \frac { \partial ^ { 2 } L ( y _ { i } , \hat { y } _ { i } ^ { ( t - 1 ) } ) } { \partial \hat { y } _ { i } ^ { ( t - 1 ) } }$$
	Since $\hat { y } _ { i } ^ { t - 1 }$ is a known value at step t, it is a constant, which has no effect on the optimization of the function, so the objective function can be further written as: $$Obj ^ { ( t ) } \approx \sum _ { i = 1 } ^ { n } [ g _ { i } f _ { t } ( x _ { i } ) + \frac { 1 } { 2 } h _ { i } f _ { t } ^ { 2 } ( x _ { i } ) ] + \sum _ { i = 1 } ^ { t } \Omega ( f _ { i })$$ So we just need to find the values of the first-order derivative and second-order derivative of the loss function at each step, and then optimize the objective function to get f(x) at each step, and finally get an overall model based on the additive model.
	\paragraph{Final simplification of the objective function based on the decision tree}
	
	Because the decision tree traverses the sample, it is actually traversing the leaf nodes. Therefore the problem can be transformed and the decision tree model will be defined as $f_{t}(x_i)=w_{q(x)}$ where $q(X)$ represents which leaf node the sample is on and $w$ denotes the weight on that leaf node. So $w_{q(x)}$ represents the values (predicted values) taken for each sample. This sample traversal can then be reduced to
	$$\left. \begin{array} { l }{ \sum _ { i = 1 } ^ { n } [ g _ { i } f _ { t } ( x _ { i } ) + \frac { 1 } { 2 } h _ { i } f _ { t } ^ { 2 } ( x _ { i } ) ] = \sum _ { i = 1 } ^ { n } [ g _ { i } w _ { q ( x _ { i } ) } + \frac { 1 } { 2 }h _ { i } w _ { q ( x _ { i } ) } ^ { 2 } ] = \sum _ { j = 1 } ^ { T } [ ( \sum _ { i \in I _ { j } } g _ { i } ) w _ { j } + \frac { 1 } { 2 } ( \sum _ { i \in I _ { j } } h _ { i } ) w _ { j } ^ { 2 }]}\end{array} \right.$$
	In decision trees, the complexity of the decision tree can be composed of the number of leaves T. The fewer the leaf nodes the simpler the model, and in addition the leaf nodes should not contain too much weight w , so the canonical term of the objective function can be defined as: $$ \Omega ( f _ { t } ) = \gamma T + \frac { 1 } { 2 } \lambda \sum _ { j = 1 } ^ { T } w _ { j } ^ { 2 }$$
	Thus, the objective function eventually becomes $$\left. \begin{array} { l }{ O b j ^ { ( t ) } \approx \sum _ { i = 1 } ^ { n } [ g _ { i } f _ { t } ( x _ { i } ) + \frac { 1 } { 2 } h _ { i } f _ { t } ^ { 2 } ( x _ { i } ) ] + \Omega ( f _ { t } ) } \\ { = \sum _ { i = 1 } ^ { n } [ g _ { i } w _ { q ( x _ { i } ) } + \frac { 1 } { 2 } h _ { i } w _ { q ( x _ { i } ) } ^ { 2 } ] + \gamma T + \frac { 1 } { 2 } \lambda }\sum_{j=1}^Tw_j^2 \\ { = \sum _ { j = 1 } ^ { T } [ ( \sum _ { i \in I _ { j } } g _ { i } ) w _ { j } + \frac { 1 } { 2 } ( \sum _ { i \in I _ { j } } h _ { i } + \lambda ) w_j^2] + \gamma T } \end{array} \right.$$ To simplify the expression, we then define: $G _ { j } = \sum _ { i \in I _ { j } } g _ { i } \quad H _ { j } = \sum _ { i \in I _ { j } } h _ { i }$, then the objective function is:
	$$O b j ^ { ( t ) } = \sum _ { j = 1 } ^ { T } [ G _ { j } w _ { j } + \frac { 1 } { 2 } ( H _ { j } + \lambda ) w _ { j } ^ { 2 } ] + \gamma T$$
	Then by taking the first order derivative of the objective function with respect to $w_j$ and making it equal to 0, the weights corresponding to the leaf node j can be found: $w _ { j } ^ { * } = - \frac { G _ { j } } { H _ { j } + \lambda }$. The objective function of the decision tree based XGBoost model is thus obtained as $$O b j = - \frac { 1 } { 2 } \sum _ { j = 1 } ^ { T } \frac { G _ { j } ^ { 2 } } { H _ { j } + \lambda } + \gamma T$$
	\paragraph{Optimal cut-point partitioning algorithm and optimization strategy}
	
	1. Enumerate all available features for each leaf node, starting from a tree of depth 0.\\
	2. For each feature, rank the training samples belonging to that node in ascending order according to the value of that feature, determine the best splitting point for that feature by a linear scan, and record the splitting gain of that feature.\\
	3. Select the feature with the largest gain as the splitting feature, use the best splitting point of that feature as the splitting location, split the left and right two new leaf nodes on that node, and associate the corresponding sample set for each new node.\\
	4. Return to step 1 and execute recursively until a specific condition is satisfied.
	\paragraph{Predict the sample values using the new decision tree and accumulate them to the original values:}
	Several decision trees are trained by addition. The so-called additive training is essentially a meta-algorithm that applies to all additive models, and it is a heuristic algorithm. Using additive training, our goal is no longer to directly optimize the entire objective function, but to optimize the objective function in steps, first optimizing the first tree, and then optimizing the second tree after that, until we have optimized K trees.
	
	\section{Experimental Process and Results Demonstration}
	\subsection{Data Processing}
	Process the data so that the final data includes time, aqi and the concentration of six pollutants ($SO_2,NO_2,CO,O_3,PM10,PM2.5$). Then visualize them:
	
	\begin{figure}[!htb]
		\centering
		\subfigure
		{
			\begin{minipage}[b]{0.23\linewidth}
				\centering
				\includegraphics[width=\hsize]{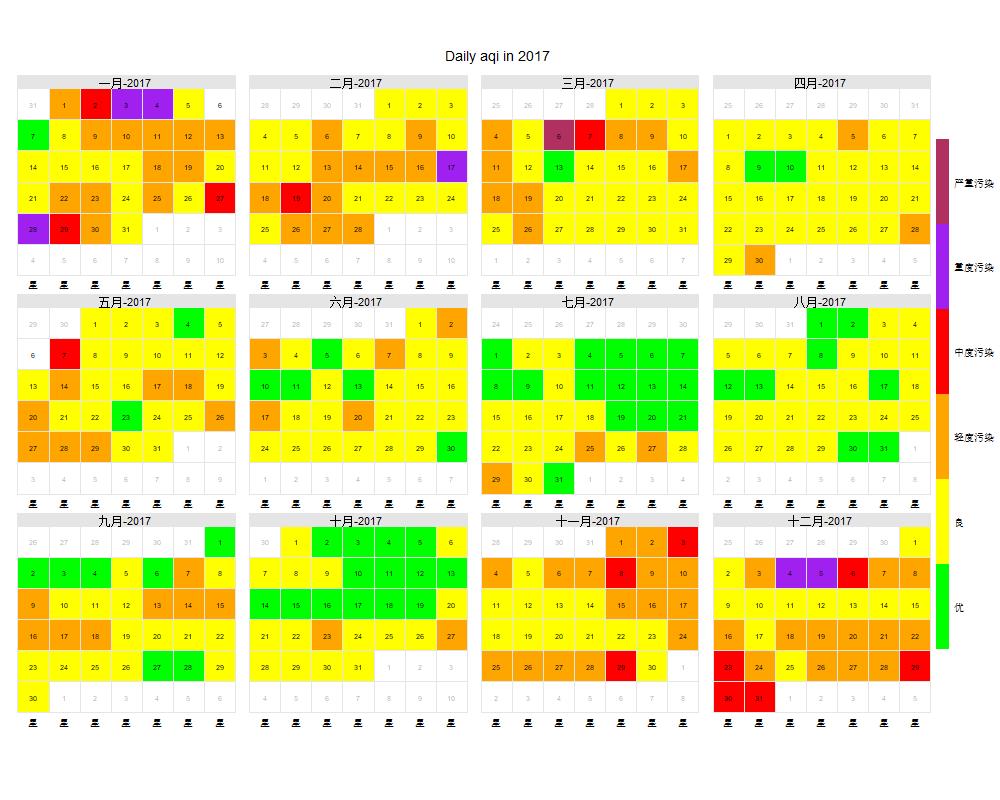}               
				\label{(a)aqi_2017}
			\end{minipage}
		}
		\subfigure
		{
			\begin{minipage}[b]{0.23\linewidth}
				\centering
				\includegraphics[width=\hsize]{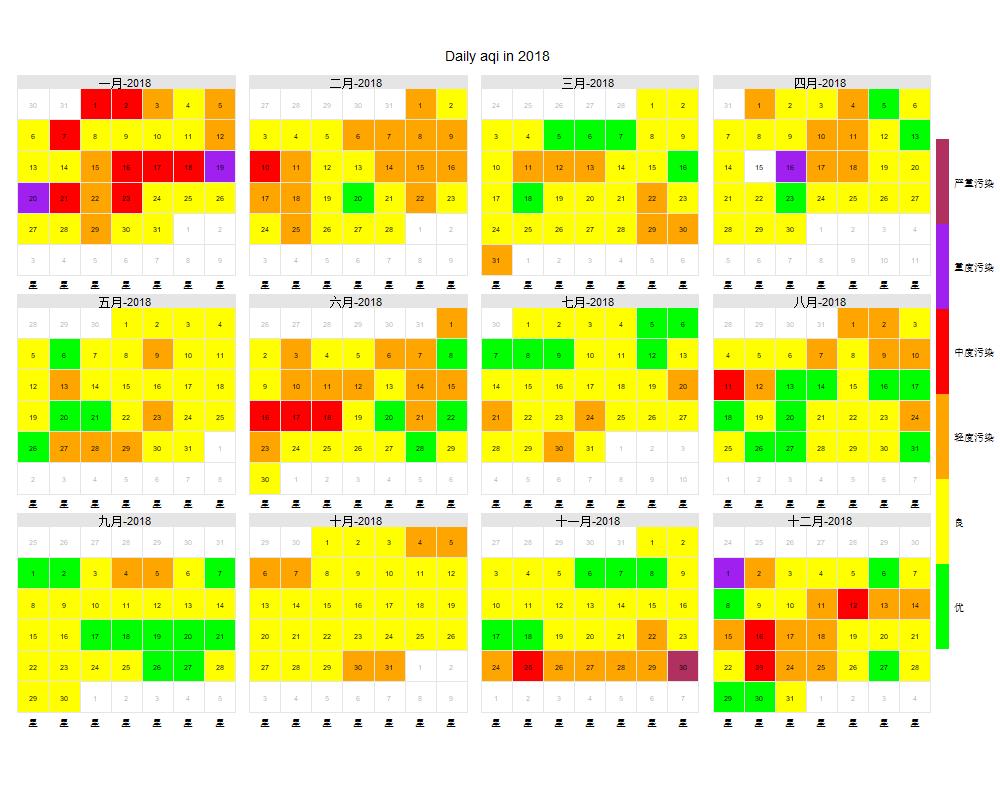}               
				\label{(b)aqi_2018}
			\end{minipage}
		}
		\subfigure
		{
			\begin{minipage}[b]{0.23\linewidth}
				\centering
				\includegraphics[width=\hsize]{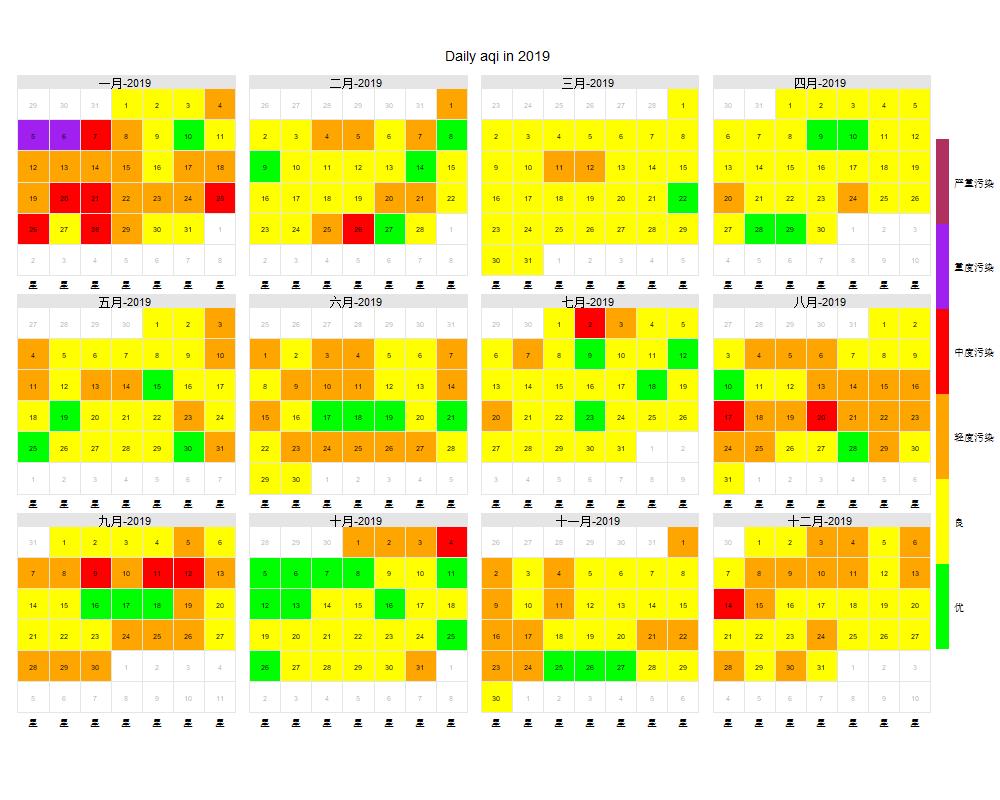}               
				\label{(c)aqi_2019}
			\end{minipage}
		}
		\subfigure
		{
			\begin{minipage}[b]{0.23\linewidth}
				\centering
				\includegraphics[width=\hsize]{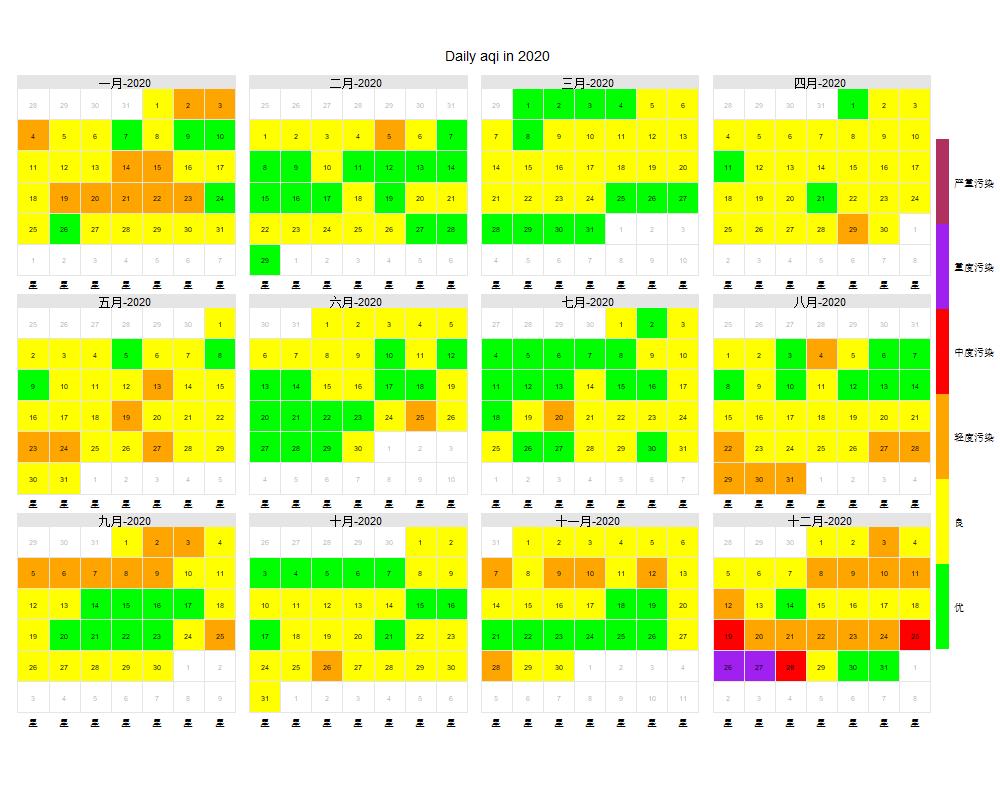}               
				\label{(d)aqi_2020}
			\end{minipage}
		}
		\caption{AQI from 2017 to 2020}
	\end{figure}
	
	\begin{figure}[!htb]
		\centering
		\subfigure
		{
			\begin{minipage}[b]{0.45\linewidth}
				\centering
				\includegraphics[width=\hsize]{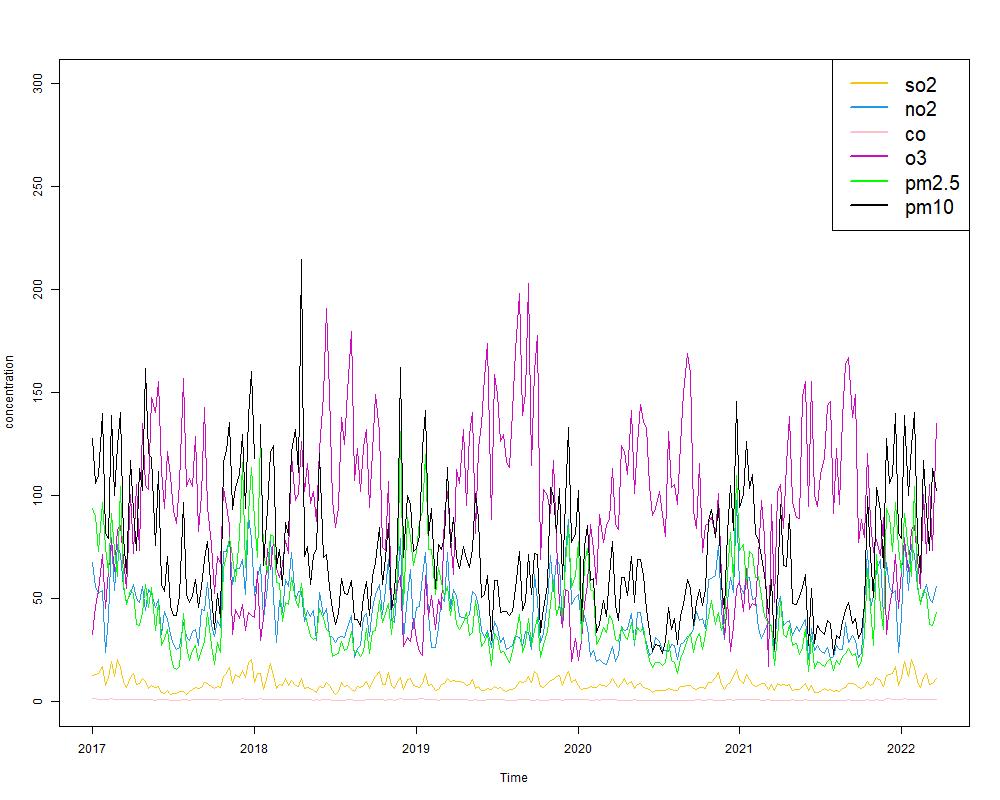}
			\end{minipage}
		}
		\subfigure
		{
			\begin{minipage}[b]{0.45\linewidth}
				\centering
				\includegraphics[width=\hsize]{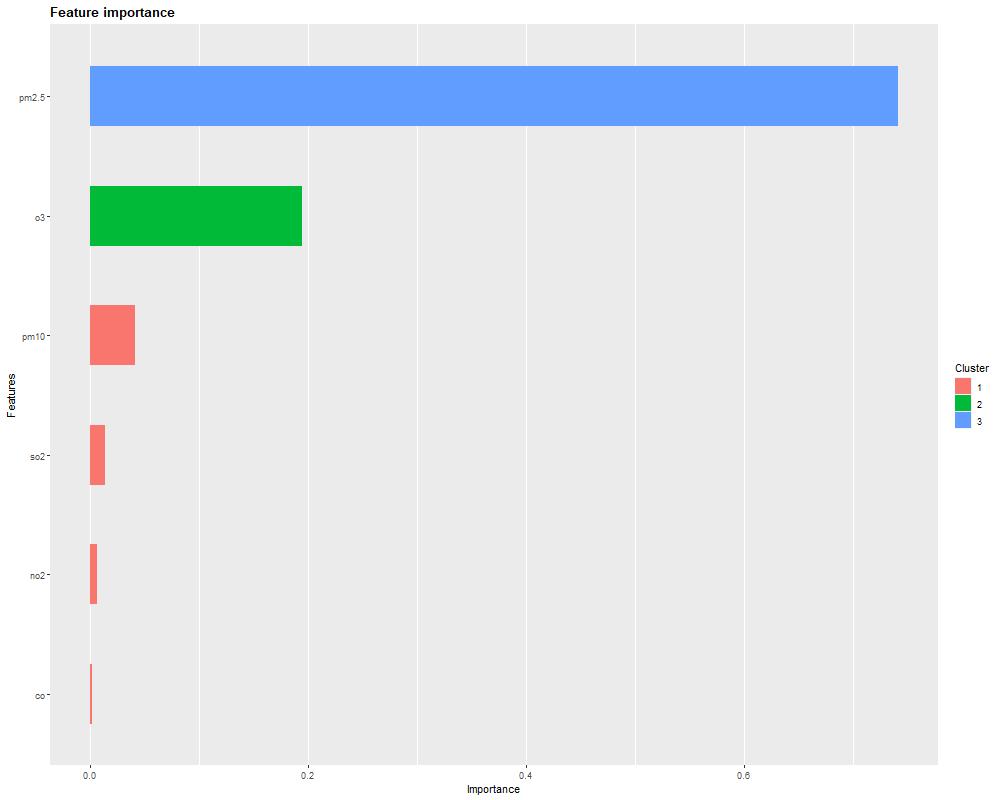}
			\end{minipage}
		}
		\caption{Time series plot and The degree of influence of six pollutants on AQI}
	\end{figure}
	
	Figure 1 visualizes the AQI from 2017-01-01 to 2020-12-31, where the darker color means higher AQI and higher air pollution level. It is obvious that the air quality in 2020 has improved significantly compared with previous years, which initially confirms our speculation that "lockdown can reduce air pollution".
	Figure 2(left panel) shows the time series of the concentrations of the six pollutants (here the weekly average series of the concentrations are plotted). It can be seen that $SO_2,NO_2,CO,PM10,PM2.5$ have similar trends, but due to the influence of seasonal factors, it is not clear that there is a significant decrease in concentrations after the lockdown. The $O_3$ concentration, on the other hand, has an opposite trend to the other pollutants.
	Figure 2(right panel) shows the degree of influence of different pollutants on AQI based on the XGBoost model by dividing the data into training and test sets and transforming AQI into a 0-1 factorial variable with 150 as the cutoff. From the figure, it can be seen that PM2.5 has the greatest degree of influence on AQI in this data set.
	\begin{figure}[!htb]
		\centering
		\subfigure
		{
			\begin{minipage}[b]{0.23\linewidth}
				\centering
				\includegraphics[width=\hsize]{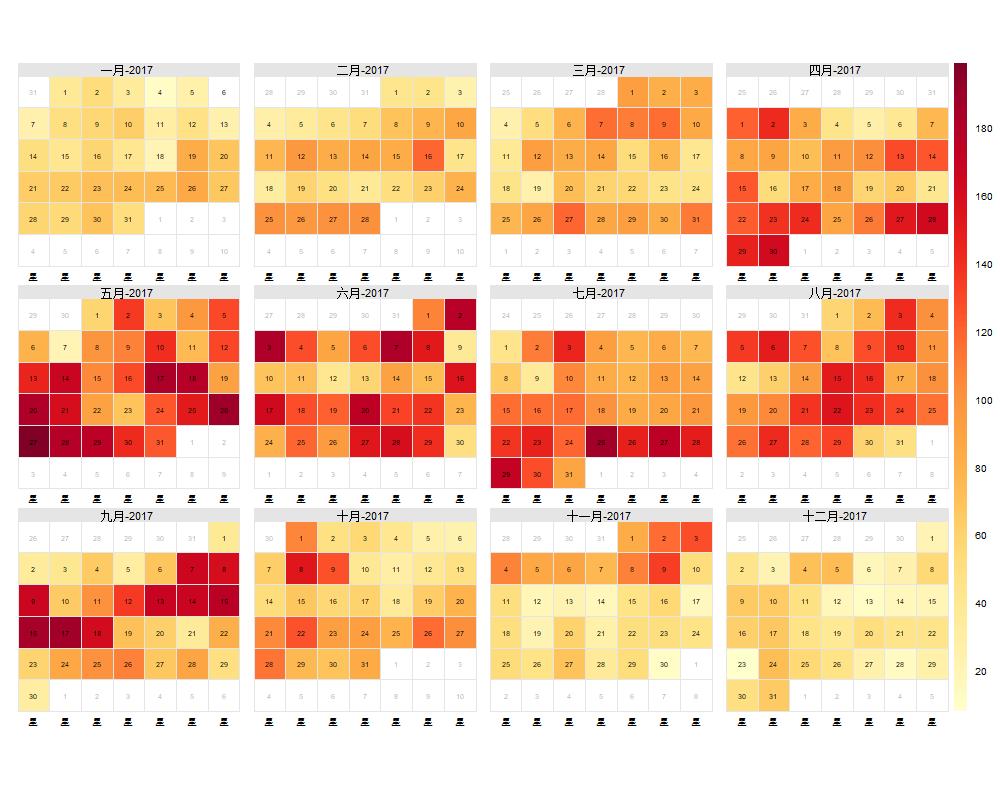}               
				\label{(a)o3_2017}
			\end{minipage}
		}
		\subfigure
		{
			\begin{minipage}[b]{0.23\linewidth}
				\centering
				\includegraphics[width=\hsize]{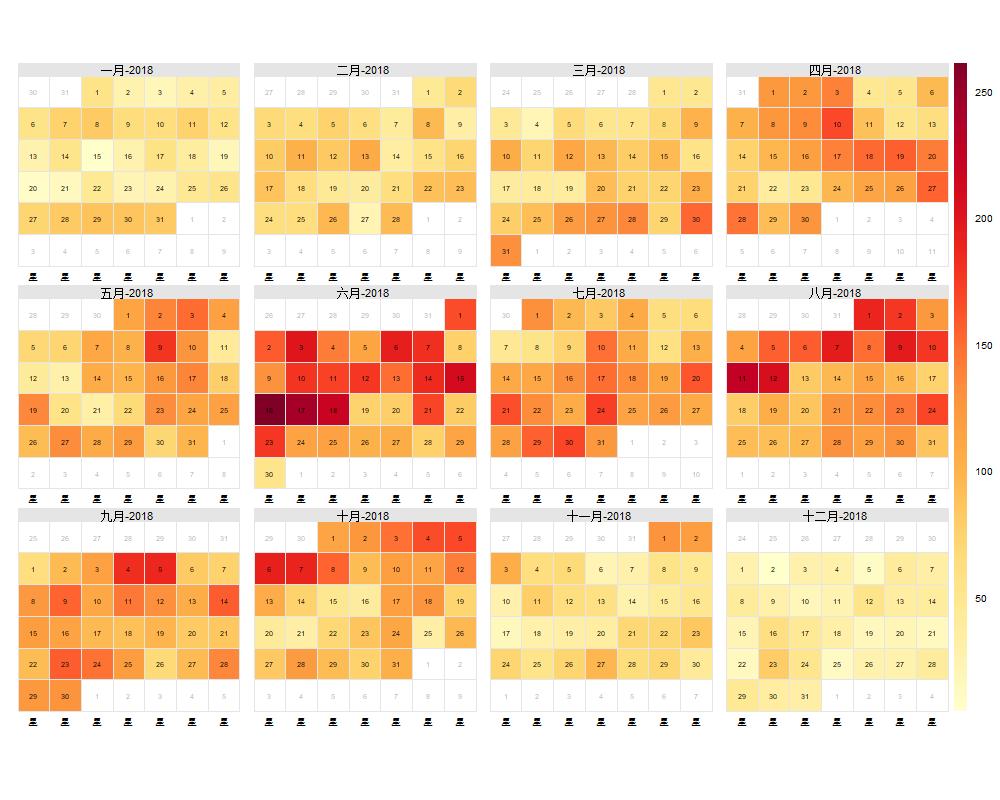}               
				\label{(b)o3_2018}
			\end{minipage}
		}
		\subfigure
		{
			\begin{minipage}[b]{0.23\linewidth}
				\centering
				\includegraphics[width=\hsize]{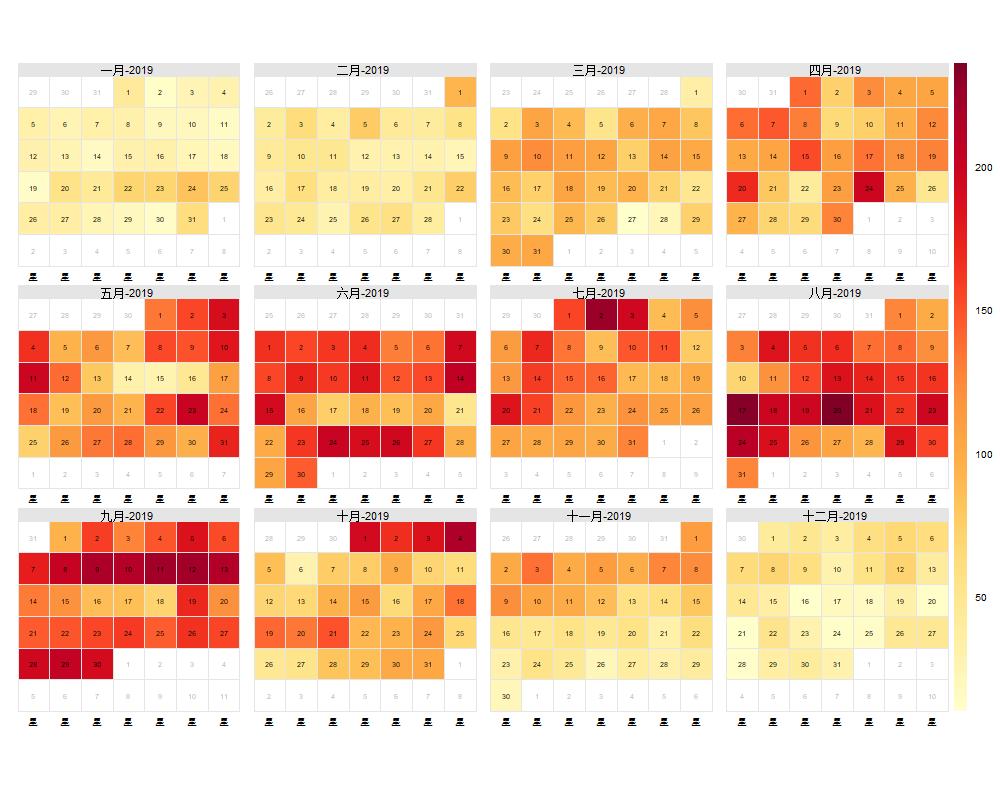}               
				\label{(c)o3_2019}
			\end{minipage}
		}
		\subfigure
		{
			\begin{minipage}[b]{0.23\linewidth}
				\centering
				\includegraphics[width=\hsize]{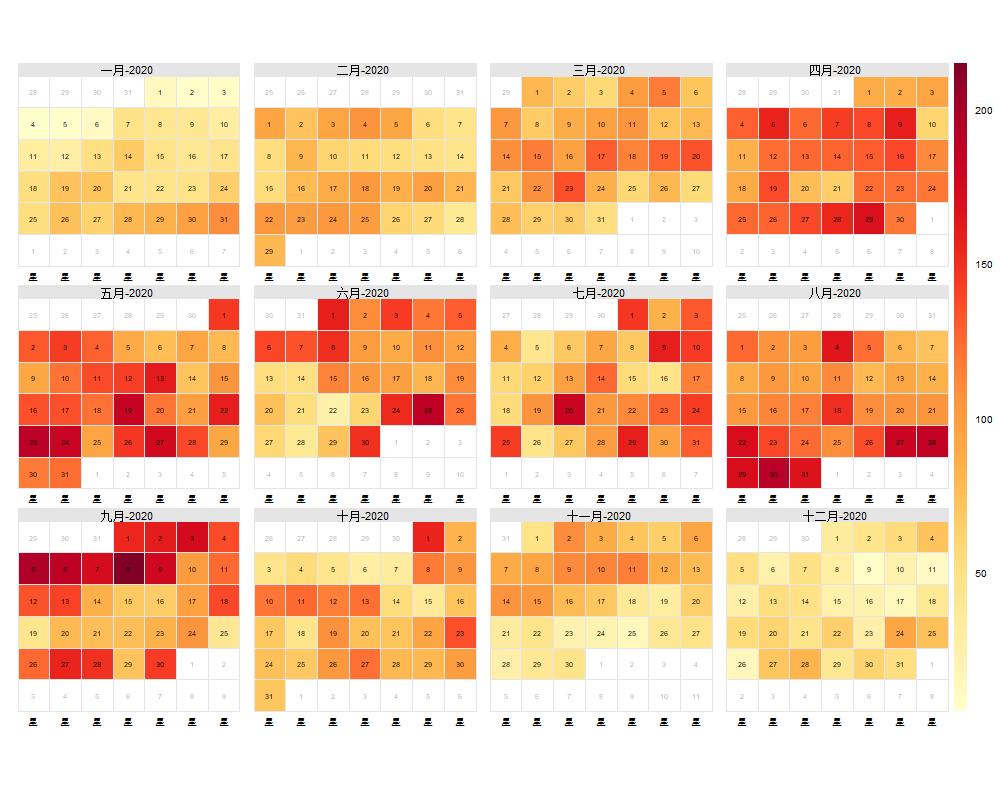}               
				\label{(d)o3_2020}
			\end{minipage}
		}
		\caption{$O_3$ from 2017 to 2020}
	\end{figure}
	
	Since in Figure 2(left panel), the $O_3$ trends are very different, Figure 3 visualizes the concentrations of $O_3$ from 2017-01-01 to 2020-12-31. It can be found that the $O_3$ concentration in 2020 has an increasing trend compared to the previous years instead. To verify our initial conjecture, the following counterfactual time series predictions will be made for $PM2.5$ (the most influential),$ NO_2 $(representative of the remaining four pollutants), and $O_3$ (with anomalous trends) respectively.

	\begin{figure}[!htb]
		\centering
		\subfigure
		{
			\begin{minipage}[b]{0.3\linewidth}
				\centering
				\includegraphics[width=\hsize]{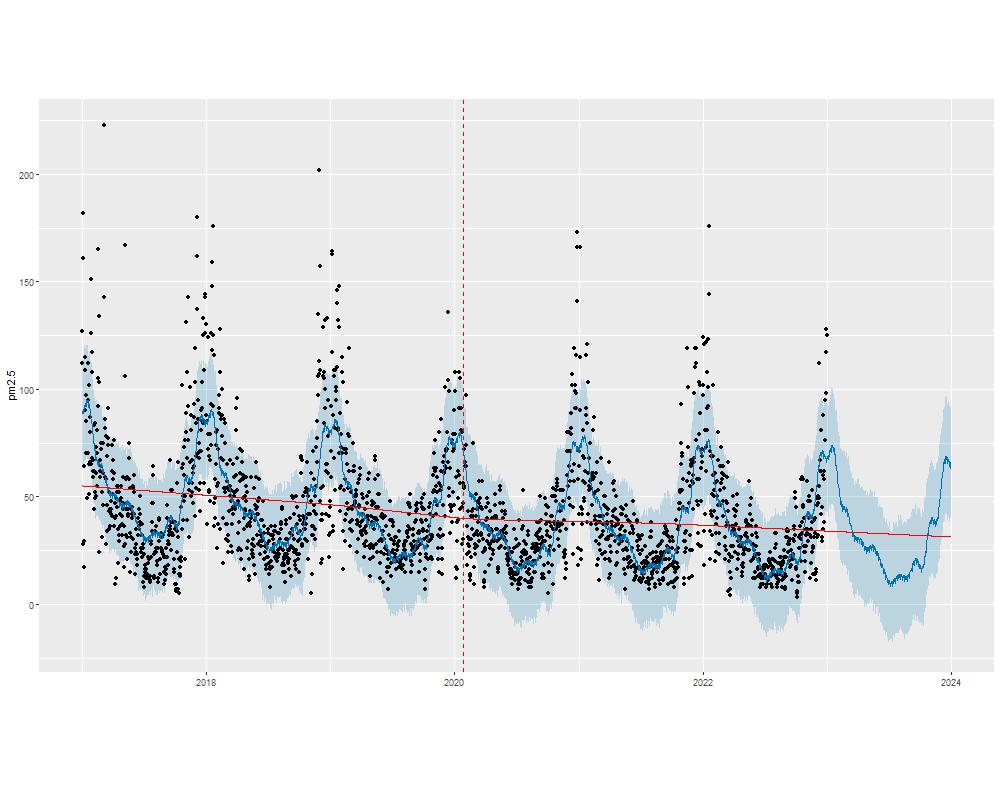}               
				\label{pm2.5}
			\end{minipage}
		}
		\subfigure
		{
			\begin{minipage}[b]{0.3\linewidth}
				\centering
				\includegraphics[width=\hsize]{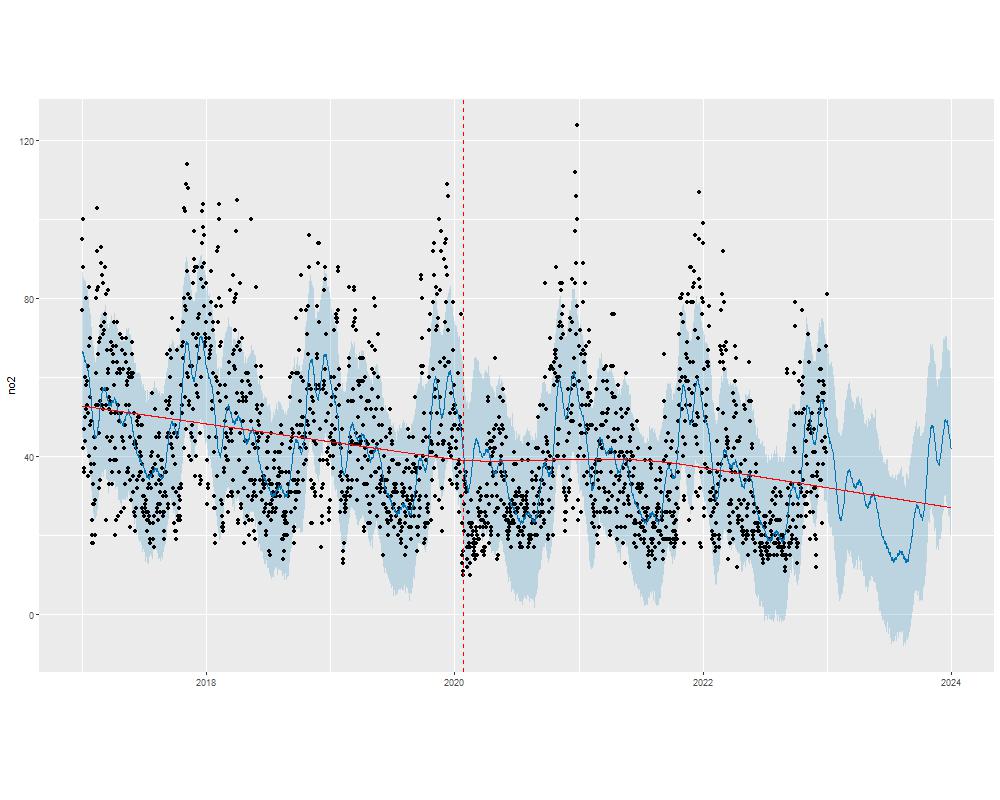}               
				\label{(b)no2_prophet}
			\end{minipage}
		}
		\subfigure
		{
			\begin{minipage}[b]{0.3\linewidth}
				\centering
				\includegraphics[width=\hsize]{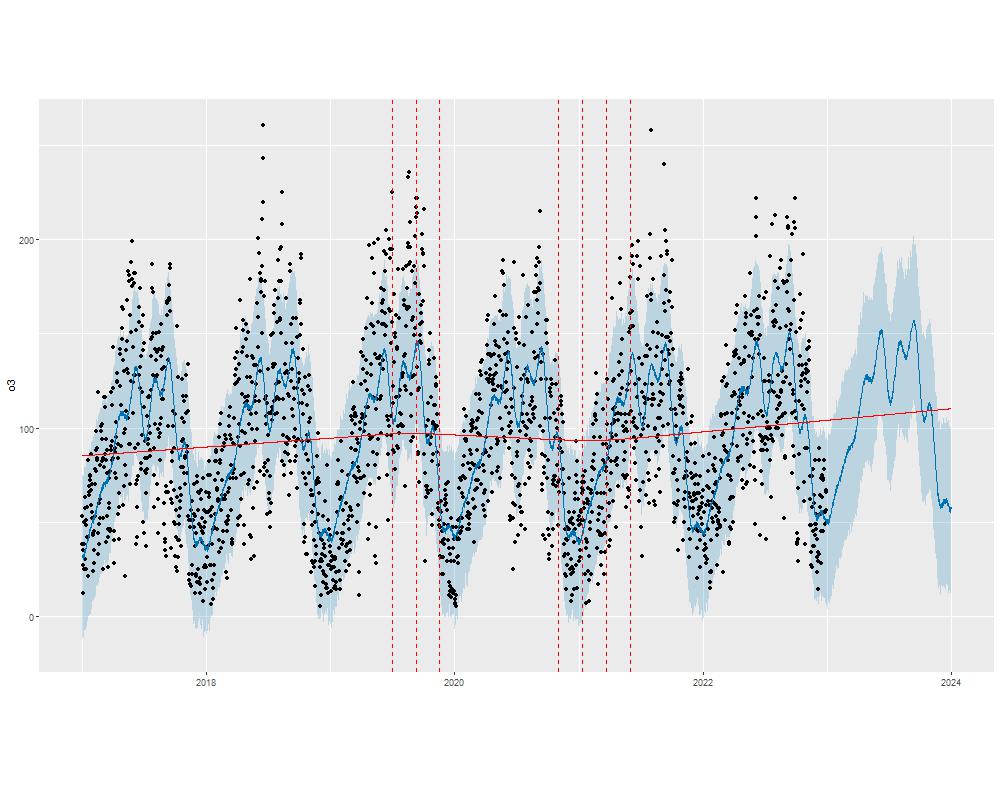}               
				\label{(c)o3_prophet}
			\end{minipage}
		}
		\caption{ prophet analysis and prediction for $PM2.5$, $ NO_2 $ and $O_3$}
	\end{figure}
	
	Then we use the Facebook prophet algorithm to view general trends and main changepoints of data. In Figure 4, red lines represent trends based on the piecewise linear functions that were decomposed from the time series sequences, and vertically dotted red lines signify the most prominent changepoints. As we can see, there are prominent changepoints at the beginning of the year 2020(2020-01-27) on $PM2.5$ and $ NO_2 $ sequences. They all show a downward trend in the following seasons, which demonstrates our conjecture perfectly. As for the $O_3$, however, it showcases an abnormal trend of generally upward after the year 2022. 
	
	\subsection{Results}
	We fit SARIMA, LSTM and XGBoost models for $NO_2$, $PM2.5$ and $O_3$ by the history data from January 01, 2017, to December 31, 2019. From the fitted models, we predict air pollution counterfactual levels (absent the pandemic) during a 4-month period from January 01, 2020, to April 30, 2020. 
	
	\subsubsection{Model assessment}
	To assess the overall predictive performance of the models above, we repeated the same procedure of model building and prediction as described above, this time training the model based on the data from January 01, 2017 to December 31, 2018, and predicting for a period from January 01, 2019 to April 31, 2019. This allows us to assess model fit and evaluate our modeling approach absent the pandemic. The main goal of implementing this assessment is to find out the model’s performance in prediction absent the pandemic and compare its predictive performance using the mean squared error.
	
	\begin{table}[]
		\caption{Mean Squared Error}
		\centering
		\begin{tabular}{llll}
			\toprule
			& SARIMA & LSTM   & XGBoost \\ 
			\midrule
			$N0_2$  & 161.19 & 160.62 & 648.68  \\ 
			$PM2.5$ & 546.49 & 581.10 & 1089.78 \\ 
			$O_3$   & 719.02 & 771.07 & 1531.79 \\ 
			\bottomrule
		\end{tabular}
	\end{table}
	
	From the MSE table, we found that the overall effect of the SARIMA model was better than that of the other two models.Reflecting, we think this is because SARIMA models are well-suited for time series data that exhibit seasonality, while LSTM and XGBoost are more general-purposemodels that can be applied to a variety of data types. ln situations where the datahas a clear seasonal component, SARIMA may be able to better capture theunderlying pattern and make more accurate predictions. Additionally, SARIMAmodels are relatively simple and interpretable, making them easier to understandand debug.
	
	We applied the above three models for each pollutant, and by calculating the mse of the prediction results of different models, the following shows the results of the model with the best fitted prediction for each pollutant(figure 5-7):
	
	\begin{figure}[!htb]
		\centering
		\subfigure
		{
			\begin{minipage}[b]{0.45\linewidth}
				\centering
				\includegraphics[width=\hsize]{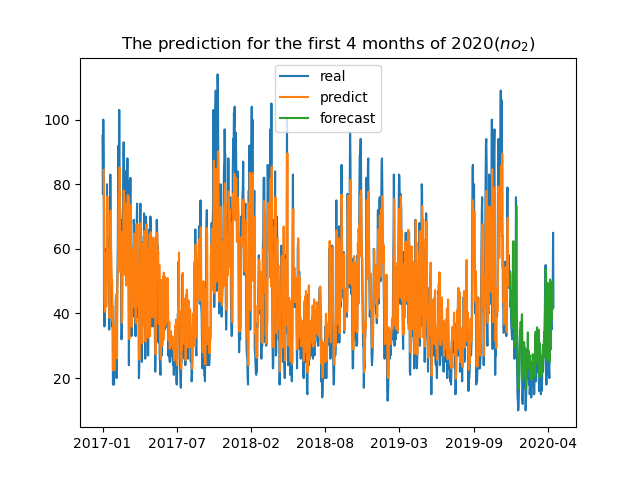}               
				\label{(a)no2_predict}
			\end{minipage}
		}
		\subfigure
		{
			\begin{minipage}[b]{0.45\linewidth}
				\centering
				\includegraphics[width=\hsize]{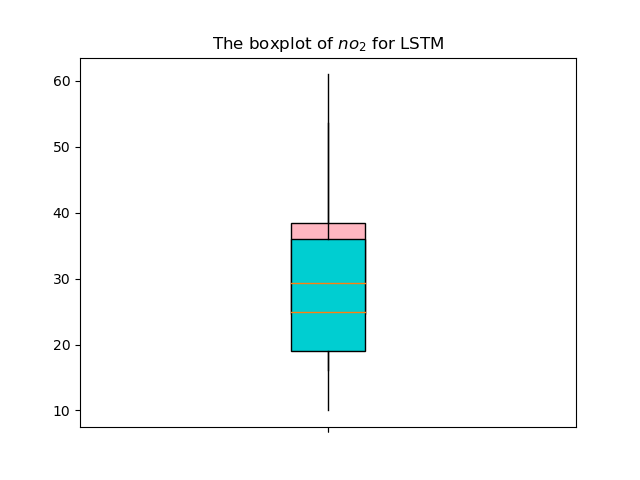}               
				\label{(b)no2_boxplot}
			\end{minipage}
		}
		
		\caption{ $ NO_2 $ LSTM}
	\end{figure}
	
	\begin{figure}[!htb]
		\centering
		\subfigure
		{
			\begin{minipage}[b]{0.45\linewidth}
				\centering
				\includegraphics[width=\hsize]{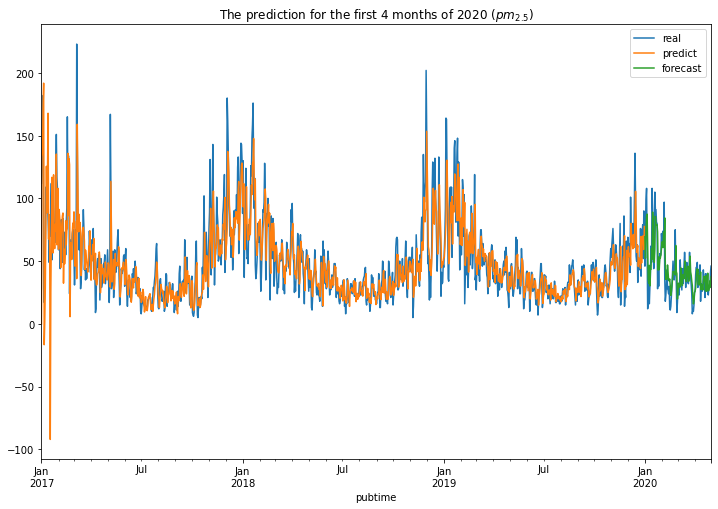}               
				\label{(a)pm2.5_predict}
			\end{minipage}
		}
		\subfigure
		{
			\begin{minipage}[b]{0.45\linewidth}
				\centering
				\includegraphics[width=\hsize]{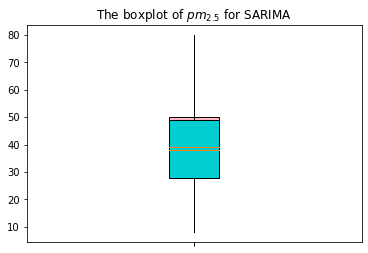}               
				\label{(b)pm2.5_boxplot}
			\end{minipage}
		}
		
		\caption{ $ PM2.5 $ SARIMA}
	\end{figure}
	
	\begin{figure}[!htb]
		\centering
		\subfigure
		{
			\begin{minipage}[b]{0.45\linewidth}
				\centering
				\includegraphics[width=\hsize]{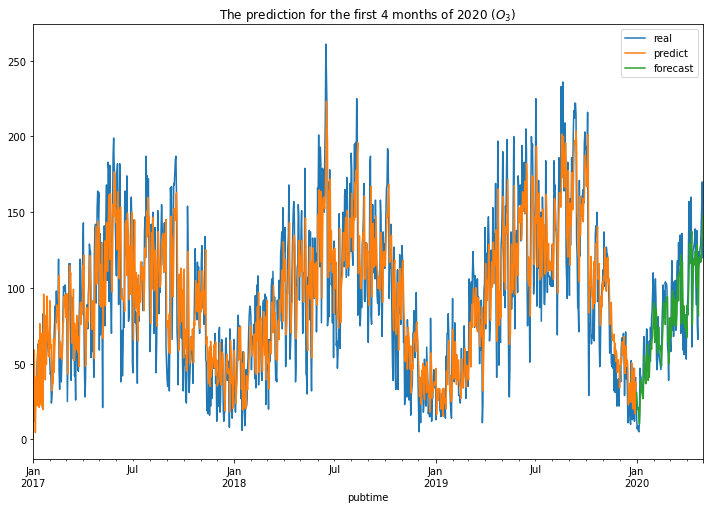}               
				\label{(a)O3_predict}
			\end{minipage}
		}
		\subfigure
		{
			\begin{minipage}[b]{0.45\linewidth}
				\centering
				\includegraphics[width=\hsize]{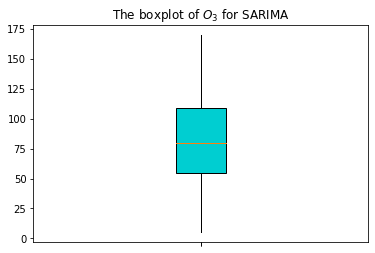}               
				\label{(b)O3_boxplot}
			\end{minipage}
		}
		
		\caption{ $ O_3 $ SARIMA}
	\end{figure}
	
	From the 3 models above, we can find the counterfactual predictions (i.e., assuming the pandemic did not occur) were significantly higher than the observed pollutant values after the outbreak of the pandemic.

	\section{Conclusion}
	Through the data visualization above, the prophet algorithm and the counterfactual predictions based on different models, we can find a significant decrease in the concentration of $NO_2$, which is due to the lockdown that reduces industrial and traffic emissions of exhaust gases, of which nitrogen oxides are the main component. The $PM2.5$ has a decreasing trend, and the prophet algorithm also considers its concentration change as anomalous in early 2020. However, the reduction of $PM2.5$ is not significant in the counterfactual prediction, which is very confusing because $PM2.5$ is also one of the main components of exhaust gas, and its decrease should be similar to that of $NO_2$. In addition, the increasing trend of $O_3$ concentration is more puzzling. Why did the $O_3$ pollution increase when the sky became blue? Therefore, we studied the data of several other cities and found that $PM2.5$ decreases to different degrees depending on the area. And $O_3$ has a clear upward trend no matter where it is. 
	\begin{figure}[!htb]
		\centering
		\subfigure
		{
			\begin{minipage}[b]{0.3\linewidth}
				\centering
				\includegraphics[width=\hsize]{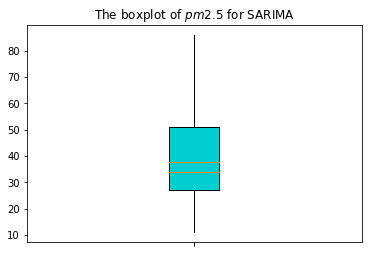}               
				\label{(a)hefei}
			\end{minipage}
		}
		\subfigure
		{
			\begin{minipage}[b]{0.3\linewidth}
				\centering
				\includegraphics[width=\hsize]{SARIMA_boxplot_pm2_5.png}               
				\label{(b)wuhan}
			\end{minipage}
		}
		\subfigure
		{
			\begin{minipage}[b]{0.3\linewidth}
				\centering
				\includegraphics[width=\hsize]{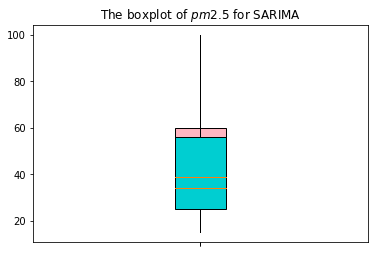}               
				\label{(c)yinchuan}
			\end{minipage}
		}
		
		\caption{ $ PM2.5 $ in HeFei, WuHan and YinChuan}
	\end{figure}

	By consulting the data, we believe that this is because: the reduction of NOx (nitrogen oxides) is significant during the lockdown. This leads to a weaker titration of NOx, which reduces the abatement of $O_3$, and therefore the atmospheric $O_3$ concentration increases. At the same time, the daytime $O_3$ production and oxidation capacity increases to a small extent, resulting in an increase in daytime $OH$ radicals and nighttime $NO_3$ radical concentrations. Thus, the enhanced oxidation capacity of the atmosphere (including the increase of ozone as well as $OH$ and $NO_3$ radicals) will promote the formation of secondary particulate matter. Different wind directions will carry these particles to different areas, which results in a significant decrease in $PM2.5$ in some areas and hazy weather in others.
	In addition to the above-mentioned decrease in nitrogen oxide concentration that makes the rate of $O_3$ decomposition lower, the increase in human activities in yards and gardens during the isolation period leads to the increase in $VOCs$ concentration and promotes the accumulation of ozone.
	In addition, some scientists speculate that the absence of hazy sunlight can penetrate the air more easily after $PM 2.5$ is reduced, providing more energy for surface ozone production.

	\section{Discussion}
	1. The data contain incomplete information, and the main work is focused on the urban context, comparing the impact of COVID-19 on the environment in a macroscopic manner without quantitative analysis. Further, the air quality impacts of blockades on traffic-related roadside environments, for example, remain unclear.
	
	2. Factors affecting the scope and intensity of atmospheric pollution include the nature of pollution sources (source intensity, source height, temperature within the source, exhaust rate, etc.), meteorological conditions (wind direction, wind speed, temperature stratification, etc.), and the nature of the surface (topographic relief, roughness, ground cover, etc.). Lacking such related data, we are unaccessible to carry out multivariate time series analysis considering the above factors, so as to draw more reasonable and accurate conclusions.
	
	3. When selecting different models for data fitting, it can be found that the SARIMA model is significantly better than the other two, in addition to the data type, it may be that we do not fully grasp the parameter selection of LSTM and XGBoost models.Therefore, the optimal model is not selected . The SARIMA parameters are selected through the grid search and the AIC criterion.
	
	\bibliographystyle{unsrt}  


\end{document}